\documentclass{emulateapj}
\usepackage{apjfonts}

\begin{document}

\title{The FMOS-COSMOS survey of star-forming galaxies at $z\sim 1.6$. I. H$\alpha$-based star formation rates and dust extinction}

\begin{abstract}

We present the first results from a near-IR spectroscopic survey of the COSMOS field, using the Fiber Multi-Object Spectrograph on the Subaru telescope, designed to characterize the star-forming galaxy population at $1.4<z<1.7$.  The high-resolution mode is implemented to detect H$\alpha$ in emission between $1.6{\rm -}1.8~\mathrm{\mu m}$ with $f_{\rm H\alpha}\gtrsim4\times10^{-17}$ erg cm$^{-2}$ s$^{-1}$.  Here, we specifically focus on 271 sBzK-selected galaxies that yield a H$\alpha$ detection thus providing a redshift and emission line luminosity to establish the relation between star formation rate and stellar mass.  With further $J$-band spectroscopy for 89 of these, the level of dust extinction is assessed by measuring the Balmer decrement using co-added spectra. We find that the extinction ($0.6\lesssim A_\mathrm{H\alpha} \lesssim 2.5$) rises with stellar mass and is elevated at high masses compared to low-redshift galaxies.  Using this subset of the spectroscopic sample, we further find that the differential extinction between stellar and nebular emission \hbox{$E_\mathrm{star}(B-V)/E_\mathrm{neb}(B-V)$} is 0.7--0.8, dissimilar to that typically seen at low redshift.  After correcting for extinction, we derive an H$\alpha$-based main sequence with a slope ($0.81\pm0.04$) and normalization similar to previous studies at these redshifts.

\end{abstract}

%---------------------------------------------------------------------
%---------------------------------------------------------------------

\author{%
D.~Kashino\altaffilmark{1},
J.~D.~Silverman\altaffilmark{2}, 
G.~Rodighiero\altaffilmark{3},
A.~Renzini\altaffilmark{4}, 
N.~Arimoto\altaffilmark{5,6}, 
E.~Daddi\altaffilmark{7}, 
S.~J.~Lilly\altaffilmark{8}, 
D.~B. Sanders\altaffilmark{9}, 
J.~Kartaltepe\altaffilmark{10}, 
H.~J.~Zahid\altaffilmark{9}, 
T.~Nagao\altaffilmark{11}, 
N.~Sugiyama\altaffilmark{1,2}, 
P.~Capak\altaffilmark{12,13}, 
C.~M.~Carollo\altaffilmark{8}, 
J.~Chu\altaffilmark{9}, 
G.~Hasinger\altaffilmark{9}, 
O.~Ilbert\altaffilmark{14}, 
M.~Kajisawa\altaffilmark{15}, 
L.~J.~Kewley\altaffilmark{9,16},
A.~M.~Koekemoer\altaffilmark{17} 
K.~Kova{\v c}\altaffilmark{8}, 
O.~ Le F{\` e}vre\altaffilmark{14}, 
D.~Masters\altaffilmark{18}, 
H.~J.~McCracken\altaffilmark{19}, 
M.~Onodera\altaffilmark{8}, 
N.~Scoville\altaffilmark{20}, 
V.~Strazzullo\altaffilmark{7},
M.~Symeonidis\altaffilmark{21}		
and
Y.~Taniguchi\altaffilmark{15}
}

\email{daichi@nagoya-u.jp}

\altaffiltext{1}{% Kashino
Division of Particle and Astrophysical Science, Graduate School of Science, Nagoya University, Nagoya, 464-8602, Japan
}
\altaffiltext{2}{% John, Sugiyama
Kavli Institute for the Physics and Mathematics of the Universe (WPI), Todai Institutes for Advanced Study, The University of Tokyo, Kashiwanoha, Kashiwa 277-8583, Japan
}
\altaffiltext{3}{% Giulia
Dipartimento di Astronomia, Universit\`{a} di Padova, vicolo dell'Osservatorio 3, I-35122 Padova, Italy
}
\altaffiltext{4}{% Alvio
INAF Osservatorio Astronomico di Padova, vicolo dell'Osservatorio 5, I-35122 Padova, Italy
}
\altaffiltext{5}{% Arimoto
National Astronomical Observatory of Japan, Subaru Telescope, 650 North A'ohoku Place, Hilo, HI 96720, USA
}
\altaffiltext{6}{% Arimoto
The Graduate University for Advanced Studies, Department of Astronomical Sciences, Osawa 2-21-1, Mitaka, Tokyo 181-8588, Japan
}
\altaffiltext{7}{% Daddi
CEA-Saclay, Service d'Astrophysique, F-91191 Gif-sur-Yvette, France
}
\altaffiltext{8}{% Lilly, Kovac,  Carollo Onodera, Schawinski
Institute for Astronomy, ETH Z\"{u}rich, Wolfgang-Pauli-strasse 27, 8093 Z\"{u}rich, Switzerland
}
\altaffiltext{9}{% Sanders, Jeyhan, Gunther, Kewley
Institute for Astronomy, University of Hawaii, 2680 Woodlawn Drive, Honolulu, HI 96822, USA
}
\altaffiltext{10}{%Jeyhan
National Optical Astronomy Observatory, 950 N. Cherry Ave., Tucson, AZ 85719, USA
}
\altaffiltext{11}{% Nagao
The Hakubi Center for Advanced Research, Kyoto University, Kyoto 606-8302, Japan
}
\altaffiltext{12}{% Capak
California Institute of Technology, 1200 E. California Blvd., Pasadena, CA 91125, USA
}
\altaffiltext{13}{% Capak
Spitzer Science Center, MS 314-6, California Institute of Technology, Pasadena, CA 91125, USA
}
\altaffiltext{14}{%
Aix Marseille Universit\'e, CNRS, LAM (Laboratoire d'Astrophysique de Marseille) UMR 7326, F-13388, Marseille, France
}
\altaffiltext{15}{% Kajisawa, Taniguchi
Research Center for Space and Cosmic Evolution, Ehime University, Bunkyo-cho 2-5, Matsuyama, Ehime 790-8577, Japan
}
\altaffiltext{16}{%Kewley
Research School of Astronomy and Astrophysics, The Australian National University, Cotter Road, Weston Creek, ACT 2611, Australia
}
\altaffiltext{17}{%Anton Koekemoer
HST and JWST Instruments/Science Division, Space Telescope Science Institute, 3700 San Martin Drive, Baltimore, MD 21218, USA
}
\altaffiltext{18}{% Masters
The Observatories of the Carnegie Institution for Science, 813 Santa Barbara Street, Pasadena, CA 91101, USA
}
\altaffiltext{19}{%McCracken
Institut d'Astrophysique de Paris, UMR7095 CNRS, Universit\'e Pierre et Marie Curie, 98 bis Boulevard Arago, F-75014 Paris, France
}
\altaffiltext{20}{%Nick
California Institute of Technology, MC 249-17, 1200 East California Boulevard, Pasadena, CA 91125, USA
}
\altaffiltext{21}{% Myrto Symeonidis
Mullard Space Science Laboratory, University College London, Holmbury St. Mary, Dorking, Surrey RH5 6NT, UK
}

%---------------------------------------------------------------------
\section{Introduction}
%---------------------------------------------------------------------
Direct measurements of stellar mass ($M_\ast$) and star formation rate (SFR) of galaxies at all redshifts provide essential information for our understanding of galaxy formation and evolution.  A wide variety of SFR indicators is currently accessible through the exploitation of a broad range of the electromagnetic spectrum \citep{ken98}.  This has enabled us to establish, up to $z\sim3$, that a tight correlation exists between these two quantities, with the bulk of star-forming galaxies clustering around a ``main Sequence'' (MS)  in the $\mathrm{SFR}{\rm -}M_\ast$ plane (e.g., \citealt{elb07,dad07,noe07,sal07,pan09,kar11,whi12,zah12}, and references therein).  However, the MS slope and dispersion differ appreciably from one study to another, likely due to sample selection, adopted SFR indicator, extinction law, and the method used to measure stellar masses.  

It is clear that evolutionary studies of the MS require consistency with redshift and in particular the use of a  well-calibrated SFR indicator.  In the local universe, the most accurate and extensively implemented SFR estimators are based on the H$\alpha$ emission line luminosity as applied to large spectroscopic surveys such as the Sloan Digital Sky Survey (SDSS; \citealt{bri04,pen10}).  However, beyond a redshift of $z\sim 0.5$, H$\alpha$ is redshifted to the near-IR and one has to rely on other SFR indicators, such as [O{\sc ii}], mid-IR, UV luminosities or $1.4~\mathrm{GHz}$ radio flux.  The ``bolometric'' infrared luminosity ($8{\rm -}1000~\mathrm{\mu m}$) is another very powerful SFR indicator, but current Herschel data are not deep enough to fully map the MS, instead identifying and characterizing outlying {\it starburst} galaxies \citep{rod11}.

Recently, efforts have been made to detect H$\alpha$ in the near-IR with narrow-band imaging or grism spectroscopy (e.g., \citealt{vil08,shi09,ly11,sob13}), but new ground-based near-IR instruments (Subaru/Fiber Multi-Object Spectrograph (FMOS), Keck/MOSFIRE, the Very Large Telescope (VLT) /KMOS) will acquire spectra for large samples of galaxies at $z > 0.5$, detecting the emission lines H$\beta$, [O{\sc iii}]$\lambda$4959, 5007, H$\alpha$, [N{\sc ii}]$\lambda$6548, 6583.  In particular, Subaru/FMOS \citep{kim10}, with its high multiplex and wide field-of-view, enables us to utilize the ``local'' SFR indicator H$\alpha$ for galaxies at $0.5 \lesssim z \lesssim 1.8$.  Previous studies with FMOS have been implemented in the low-resolution mode ($R\sim 600$; e.g., \citealt{yab12,ros12}).

In the present study, we exploit the first opportunity to use FMOS in the high-resolution mode ($R\sim2600$) to detect H$\alpha$ from galaxies in the COSMOS field at $1.4 \lesssim z\lesssim 1.7$ as offered by our ``Intensive Program'' (S12B-045, PI: J.~Silverman).  Throughout this work, we assume $H_0 = 70~\mathrm{km~s^{-1}}$, $\Omega_\Lambda = 0.75$, $\Omega_\mathrm{M}=0.25$, AB magnitudes, and a \cite{sal55} initial mass function used consistently, including comparisons to published studies.

%---------------------------------------------------------------------
\section{Target selection, observations and analysis}
%---------------------------------------------------------------------
\label{sec:observations}

Our sample is chosen to represent the star-forming galaxy population in COSMOS for which we can detect H$\alpha$.  This is achieved by a selection based on stellar mass ($M_\ast > 10^{10}~M_\odot$), photometric redshift ($1.4 \lesssim  z_\mathrm{phot}\lesssim1.7$), color ($B-z$, $z-K$) and predicted H$\alpha$ flux (see below).  These are sBzK-selected \citep{dad04} galaxies, from the catalog of \cite{mcc10}, based on deep near-IR imaging from the Canada-France-Hawaii Telescope with $K_\mathrm{s}<23$, and optical imaging ($B_J$, $z^+$) from Subaru.  Photometric redshift estimates are from \citet{ilb09} based on photometry as described in \citet{cap07}.    

A critical component of our selection further depends on the SFRs and extinction properties that likely result in the detection of H$\alpha$ with a flux greater than $4\times10^{-17}~\mathrm{erg~cm^{-2}~s^{-1}}$, the limit corresponding to a line detection with a significance greater than 3$\sigma$ for integration times of 5 hr.  SFRs are estimated from the rest-frame UV luminosity as derived from $B_J$ photometry and corrected for reddening using the $B_J-z$ color (following \citealt{dad04}, 2007).  To estimate the expected H$\alpha$ flux, following \cite{cal00} we use a multiplicative factor to compensate for the differential reddening between stars and line-emitting regions ($E_\mathrm{neb}(B-V) = E_\mathrm{star}(B-V)/0.44$).  This relative reddening of stars and nebular regions is further addressed in Sections \ref{sec:BD} and \ref{sec:SFRs}.  For this study, we isolate those galaxies with an error on $E_\mathrm{star}(B-V)$ less than 0.03 mag.  We highlight that the imposition of a limit on the expected H$\alpha$ flux, at the level given above, results in a sample of target galaxies having moderately higher SFRs and lower levels of extinction ($E_\mathrm{star}(B-V)\lesssim0.5$).

We observed 755 sBzK galaxies satisfying the above criteria over six nights in 2012 March and two nights in 2013 January that essentially cover the full central square degree of COSMOS.  The $H$-long grating spans a wavelength range of $1.60\rm{-}1.80~\mathrm{\mu m}$ and has sufficient throughput to detect H$\alpha$ at these redshifts.  All data are reduced with FMOS Image based reduction package (FIBRE-pac; \citealt{iwa12}).

Spectroscopic redshifts are measured with a quality flag which describes the significance of the emission line as judged by eye in two-dimensional spectra.  A flag of 2 is indicative of a strong line covering a number of contiguous pixels with both H$\alpha$ and [N{\sc ii}]$\lambda$6583 seen is some cases.  Emission lines with low signal-to-noise (S/N), whose reliability is questionable, are assigned a flag of 1.  Of the 271 galaxies with spectroscopic redshifts, there are 168 with quality ($\mathrm{flag}=2$) line detections.  We note that our failures ($\mathrm{flag}=0$) are likely due to either an H$\alpha$ line being masked out by the OH suppression system that covers $\sim30\%{\rm -}40\%$ of the spectral range, the true redshift being outside the observed wavelength interval, or H$\alpha$ being extremely weak, possibly due to high extinction.  The latter is likely to place additional constraints on the characteristics of the observed sample.

We acquired additional spectroscopy using the $J$-long grating ($1.11{\rm -}1.35~\mathrm{\mu m}$; $R\sim1900$) to detect the H$\beta$ (and [O{\sc iii}]$\lambda$5007) required to measure the Balmer decrement and correct for extinction.  These observations were carried out in 2012 March, 2012 December and 2013 February through a cooperative effort with the University of Hawaii (PI: D.~Sanders).  In our sample, 182 galaxies have both $H$- and $J$-long coverage and 89 of them have a quality detection of H$\alpha$.  While most individual spectra do not show a clear detection of H$\beta$, the emission line is highly significant in the mean spectra (see below).  We refer the reader to a dedicated paper (J.~D.~Silverman et al., in preparation) that fully describes the survey including target selection (with a departure from color selection), redshift success rates, and inherent biases.    
 
% Figure 1
\begin{figure*}[t]
	\centering
	\includegraphics[width=5.5in]{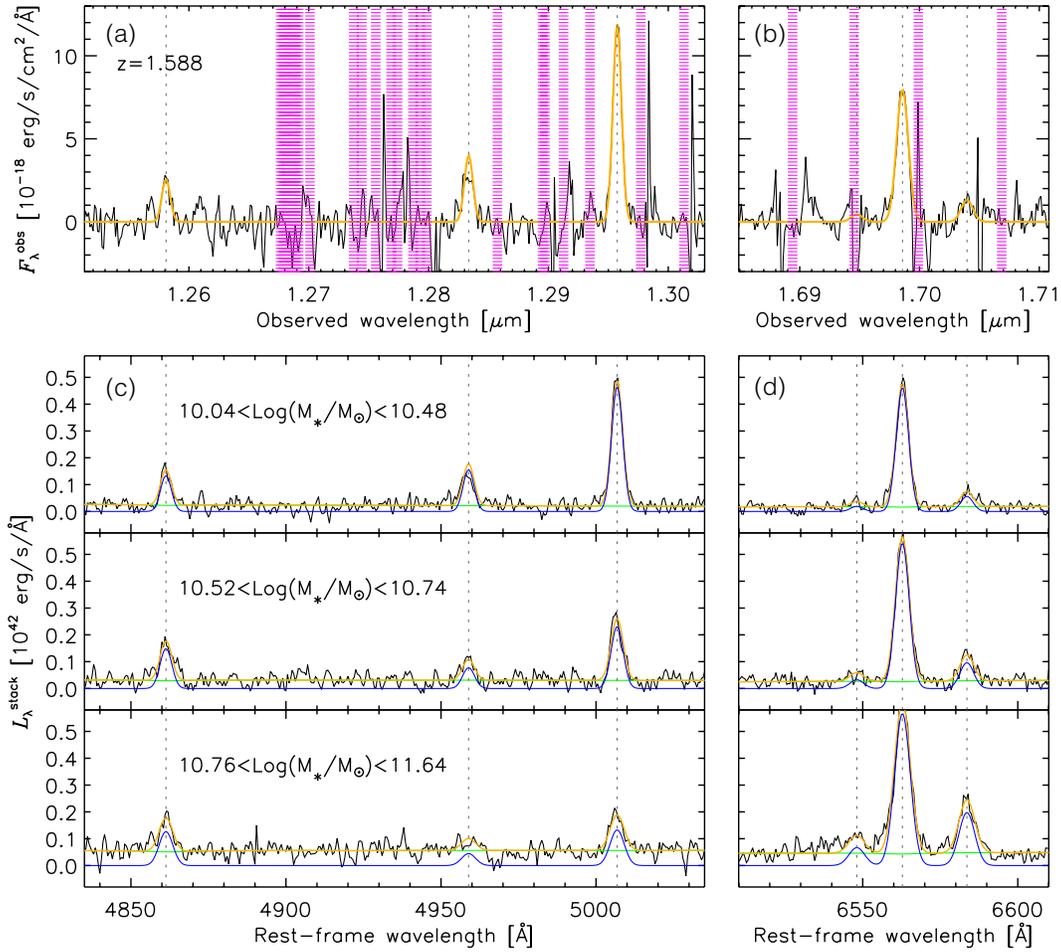} 
	\caption{Top panels: example spectrum of a galaxy at $z=1.588$ with strong emission lines detected using both the $J$-long ((a); H$\beta$, [O{\sc iii}]$\lambda \lambda 4959,~5007$) and $H$-long ((b); H$\alpha$, [N{\sc ii}]$\lambda\lambda6548,6583$) gratings.  Spectral intervals effected by OH emission are shown (hatched regions).  The emission line model is shown in yellow. Bottom panels: mean spectrum in three mass bins.  Fits to the mean spectrum are indicated by the green (continuum), blue (emission lines), and yellow (total) curves.}
	\label{fig:spec}
\end{figure*}

%---------------------------------------------------------------------
\section{Spectral analysis}
%---------------------------------------------------------------------
\subsection{Emission Line Fitting}

We run a fitting algorithm to measure the flux and associated error of emission lines observed in the $H$ (H$\alpha$ and [N{\sc ii}]$\lambda\lambda 6583,6548$) and $J$ (H$\beta$ and [O{\sc iii}]$\lambda\lambda 5007, 4959$) bands.  To mitigate the impact of residual OH features and account for errors, we define a weighting function as follows:
\begin{equation}
	W (\lambda) = \left\{ \begin{array}{ll} 
	1/N^2 (\lambda) &  \mbox{(clean window)}\\
	0 &   \mbox{(impacted by OH airglow)}
	\end{array}
	\right.
	\label{eq:weight}
\end{equation}
where $N (\lambda)$ is an error spectrum.  Each emission line is modeled by a Gaussian with the lower bound on the FWHM set to the spectral resolution of the instrument ($\sim 120$ and $90~\mathrm{km~s^{-1}}$ for $J$-long and $H$-long respectively).  We fix the width of all lines to be equivalent and the flux ratios $\mbox{[N{\sc ii}]}\lambda 6583/\mbox{[N{\sc ii}]}\lambda6548$ and $\mbox{[O{\sc iii}]}\lambda 5007/\mbox{[O{\sc iii}]}\lambda4959$ to the theoretical values of 2.96 and 2.98 \citep{sto00}, respectively.  In Figures~\ref{fig:spec}(a) and (b), we show an example spectrum with a best-fit model.  

We account for flux falling outside the aperture of the FMOS fiber of $1.\!\!\arcsec2$ diameter by the seeing conditions and the extended nature of our sources.  For each galaxy we estimate the fraction of the total light sampled by fibers by using the {\it Hubble Space Telescope}\//Advanced Camera for Surveys (ACS) $I_\mathrm{F814W}$-band images \citep{koe07}, this amounts to an assumption that the rest-frame UV and H$\alpha$ emission have the same spatial distribution.  This is justified by the tight correlation between the half-light radii in H$\alpha$  and the ACS $I$-band for the SINS/zCSINF galaxies (Mancini et al. 2011; C. Mancini and N. M. F{\" o}rster-Schreiber, private communication).  We smooth the ACS image by convolving with a point-spread function of seeing sizes of $0.8 {\rm -} 1.\!\!\arcsec2$ and then perform photometry with {\it SExtractor} \citep{ber96}.  The resulting aperture correction factors are distributed in the $\sim1.2{\rm -}5$ range, with a typical value of $\sim2$.  
For bright galaxies ($J,H < 23$) the continuum level from the flux- and aperture-corrected FMOS spectra agree with the $J$-band and  $H$-band photometry. 

\subsection{Co-added Spectra}
\label{sec:stack}
Since our data do not permit an evaluation of the Balmer decrement for most individual galaxies, we measure the average ratio of $\mathrm{H\alpha/H\beta}$ as a function of both stellar mass and reddening $E_\mathrm{star}(B-V)$.  With 89 galaxies having both a quality H$\alpha$ detection and a $J$-band spectrum, we split galaxies into three mass and reddening bins and average their de-redshifted spectra. We perform a simple average using only pixels free of OH emission, without error weighting spectra to avoid biasing the stack toward lower extinction, typical of galaxies with high S/N line detections.  In Figures \ref{fig:spec}(c) and (d), we show the mean spectra binned by stellar mass.

%Figure 2
 \begin{figure*}[t]
 	\centering
         \includegraphics[height=2.5in]{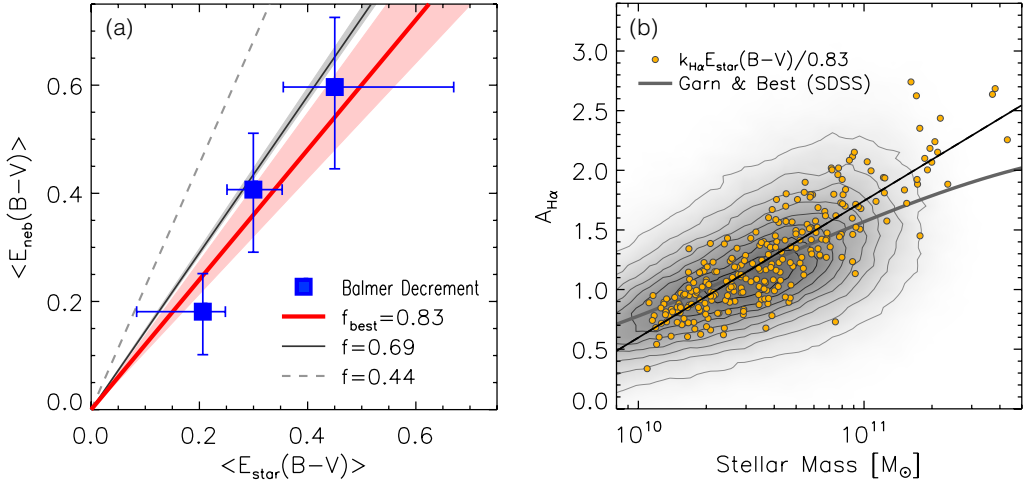}
\caption{(a) $\left<E_\mathrm{neb}(B-V)\right>$ vs. $\left<E_\mathrm{star}(B-V)\right>$.  Filled squares indicate the results from our co-added spectra with a best-fit relation (thick red line).  The vertical bars show the $1\sigma$ errors while the horizontal bars indicate the bin size in $E_\mathrm{star}(B-V)$.  The shaded area shows the errors on the fit.  (b) $A_\mathrm{H\alpha}$ vs. $M_\ast$: yellow circles represent the values of $k_{\rm H\alpha}E_\mathrm{star}(B-V)/0.83$ for FMOS galaxies with a best-fit line (thin solid line).  The contours in gray scale show the distribution of $\sim 191,600$ SDSS galaxies with the best-fit (thick gray line).
	}
	\label{fig:BD}
\end{figure*}

\section{Results}
% RESULT 1
\subsection{Balmer Decrement as a Dust Extinction Probe}
\label{sec:BD}

To achieve an intrinsic H$\alpha$ luminosity used for determining an extinction-free SFR, we must assess the level of attenuation.  Throughout our analysis, we assume the applicability of the \cite{cal00} extinction law ($R_V\equiv {A_V/E(B-V)}=4.05$, $k_\mathrm{H\alpha}=3.325,~k_\mathrm{H\beta}=4.598,~k_{1500}=10.33$).  The Balmer decrements  $\mathrm{H\alpha/H\beta}$ have been widely used to quantify the reddening in not only local but higher (up to $z\sim 1.5$) redshift galaxies (e.g., \citealt{ly12,dom13,mom13}):
\begin{eqnarray}
	E_\mathrm{neb} \left( B-V\right) = \frac{2.5}{k_\mathrm{H\beta}-k_\mathrm{H\alpha}} \log{ \left[ \frac{\mathrm{H\alpha/H\beta}}{2.86} \right]},
	\label{eq:BD}
\end{eqnarray}
which is applicable for the interstellar medium under the assumption of case B recombination \citep{ost06} with a gas temperature $T=10^4~\mathrm{K}$ and an electron density $n_\mathrm{e}=10^2~\mathrm{cm^{-3}}$.  While we have a measure of the color excess $E_\mathrm{star}(B-V)$ based on the observed $B_J-z$ color for each galaxy (see \citealt{dad07}), we need to establish whether this information can be used to accurately assess the extinction affecting the nebular emission.  Our aim is to establish a method to apply an extinction correction for individual H$\alpha$-detected galaxies lacking the detection of H$\beta$.

\cite{cal00} formulated the empirical relation $E_\mathrm{neb}(B-V)=E_\mathrm{star}(B-V)/f$ with $f=0.44$ by studying a sample of low-redshift starburst galaxies (see also \citealt{mou06}).  However, it is not established whether this conversion factor is applicable for star-forming galaxies at higher redshift.  While some studies at high-$z$ support the validity of this relation \citep{for09,wuy11,man11}, \cite{red10} argue that such excess reddening for nebular emission may not be needed for UV-selected galaxies at $z\sim2$ (see also \citealt{ono10}, for a similar conclusion).  Values of the $f$ factor larger than 0.44 are also favored in a recent Herschel work (M.~Pannella et al., in preparation).  Rather than being dependent on redshift, this factor may be related to specific SFR \citep[see Equations (15) and (16) of][]{wild11}.

For this exercise, we resort to using stacked spectra, as described in Section \ref{sec:stack}.  We specifically measure $\left<E_\mathrm{neb}(B-V)\right>$ in three bins of $E_\mathrm{star}(B-V)$.  The measured Balmer line fluxes require a correction for the underlying stellar absorption, which we assume  to be EW$^{{\rm H}\beta}_\mathrm{abs}=2.0~{\rm \AA}$, EW$^{{\rm H}\alpha}_\mathrm{abs}=1.8~{\rm \AA}$ \citep{nak04}.  The stellar absorption impacts the H$\beta$ luminosity at the $\lesssim 10$\% level while this is less for H$\alpha$ ($\lesssim 2$\%). As shown in Figure~\ref{fig:BD}(a),  the nebular extinction $\left<E_\mathrm{neb}(B-V)\right>$ ranges from 0.1 to 0.7 that corresponds to $0.7 \lesssim A_\mathrm{H\alpha}\lesssim2$.  Furthermore, there is a clear correspondence between the nebular and stellar extinction that can be fit with a linear relation characterized by $f=0.83\pm0.10$, a factor that is dissimilar to the canonical value of 0.44 and is consistent within the errors with an extrapolation of the relation from \cite{wild11} .

In Figure 2(b), we compare the level of extinction of FMOS and low-redshift galaxies, i.e., $A_\mathrm{H\alpha}$ (at a given stellar mass) as derived from the individual values of  $E_\mathrm{star}(B-V)$ for each galaxy and modified  by our new factor $f=0.83$.  Star-forming galaxies at $z\sim0.1$ from SDSS DR9 are indicated with stellar masses provided by MPA-JHU and the best-fit relation of \cite{gar10}.  While showing very good agreement between the two samples at the low-mass end \hbox{($M_{\ast}\lesssim6\times10^{10}$ $M_\odot$)}, the FMOS sample is elevated from the best-fit relation of SDSS galaxies at higher masses.  This is in slight disagreement with results from \citet{sob12} at $z=1.47$, who implemented a hybrid approach using [O{\sc ii}]/H$\alpha$ calibrated against the Balmer decrement of low-redshift galaxies. This inconsistency may not be surprising since the use of the [O{\sc ii}]/H$\alpha$ ratio as a indicator of dust attenuation is questionable, given that the gas-phase metallicity and the ionization parameter can also affect this ratio.  However, we cannot definitely rule out agreement with both of these studies due to the uncertainties of our extinction corrections.  A best-fit relation to the FMOS data (small yellow circles in Figure~\ref{fig:BD}(b)) can be expressed as follows:
\begin{eqnarray}
	A_\mathrm{H\alpha} = (0.60{\pm0.03}) + (1.15{\pm0.04}) \log{\left[\frac{M_\ast}{10^{10}M_\odot} \right]}.
	\label{eq:Ahalpha}
\end{eqnarray} 
In Table~\ref{tb:stack}, we list our measurements based on our line-fitting routine and provide the derived measure of extinction in bins of $E_\mathrm{star}(B-V)$ and stellar mass.

%Table 1
\begin{table*}[htbp]
\scriptsize
	\begin{center}
		\caption{Measurements Based on Co-added Spectra.}
		\begin{tabular}{lcccccccccc} 
		\hline \hline
		No.~of & \multicolumn{3}{c}{Color Excess Bin}  &~~~& \multicolumn{2}{c}{$L^\mathrm{stack}_\mathrm{obs}~\mathrm{(10^{42}~erg~s^{-1})}$} &~~~& \multicolumn{2}{c}{EW (\AA)\tablenotemark{a} } & Color Excess\tablenotemark{b}  \\
		\cline{2-4}\cline{6-7}\cline{9-10}
		 Spectra   & Min  & $\left< E_\mathrm{star}(B-V)\right>$ & Max &&  $\mathrm{H\alpha}$ & $\mathrm{H\beta}$ &&$\mathrm{H\alpha}$ & $\mathrm{H\beta}$ & $E_\mathrm{neb}(B-V)$  \\
		\hline
		30 & 0.084 & 0.2063  & 0.248 && $2.891\pm0.085$ & $0.818\pm0.059$ && $124\pm4$ & $28\pm2$ & $0.18\pm0.08$ \\
		29 & 0.251 & 0.2997 & 0.353 && $2.651\pm0.087$ & $0.576\pm0.068$ && $100\pm4$ & $18\pm2$ & $0.41\pm0.11$ \\
		30 & 0.355 & 0.4501 & 0.670 && $3.341\pm0.120$ & $0.581\pm0.090$ && $77\pm3$ & $11\pm2$   & $0.60\pm0.14$ \\
		\hline
		No.~of & \multicolumn{3}{c}{Stellar Mass Bin}  && \multicolumn{2}{c}{$L^\mathrm{stack}_\mathrm{obs}~\mathrm{(10^{42}~erg~s^{-1})}$} && \multicolumn{2}{c}{EW (\AA)\tablenotemark{a}} & Attenuation\tablenotemark{b}  \\
				\cline{2-4}\cline{6-7}\cline{9-10}
		 Spectra   & $M_\ast^\mathrm{min}$  & $\left< M_\ast \right>$ & $M_\ast^\mathrm{max}$ &&  $\mathrm{H\alpha}$ & $\mathrm{H\beta}$ &&$\mathrm{H\alpha}$ & $\mathrm{H\beta}$ & $A_\mathrm{H\alpha}$ \\
		\hline
		30 & 10.039 & 10.281  & 10.477 && $2.312\pm0.067$ & $0.583\pm0.051$ && $127\pm5$ & $23\pm2$ & $0.93\pm0.28$ \\
		29 & 10.521 & 10.611  & 10.739 && $2.999\pm0.090$ & $0.709\pm0.080$ && $107\pm4$ & $22\pm3$ & $1.11\pm0.34$ \\
		30 & 10.761 & 11.025  & 11.635 && $3.604\pm0.122$ & $0.701\pm0.095$ && $76\pm3$   & $13\pm2$   & $1.66\pm0.41$ \\
		\hline 
		\end{tabular}
	\label{tb:stack}
	\tablenotetext{1}{Equivalent widths (rest-frame) of H$\alpha$ and H$\beta$ emission line.}
	\tablenotetext{2}{Extinction derived from the Balmer decrement.}
	\end{center}
\end{table*}

% RESULT 2
\subsection{Comparing H$\alpha$- and UV-based SFR Indicators}
\label{sec:SFRs}

\begin{figure}[htbp]
	\centering
	\includegraphics[height=2.5in]{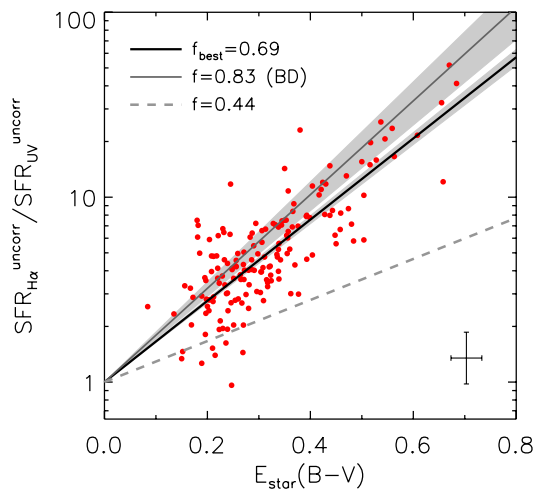} 
	\caption{Ratio of observed (not corrected for extinction) H$\alpha$- to UV-based SFRs as a function of $E_\mathrm{star}(B-V)$.  The typical errors are shown in the lower right corner.  The lines show the relations with different values of $f$ (Equation (\ref{eq:HaUVsfr})).}
	\label{fig:HaUVsfr}
\end{figure}

We now compare the SFRs derived from H$\alpha$ and from UV, assuming that both intrinsic SFRs are the same and then determine what level of extinction (as parameterized by our factor $f$) is needed to match the observed values.  In Figure \ref{fig:HaUVsfr}, we plot the ratio $\mathrm{SFR}^\mathrm{uncorr}_{\mathrm{H\alpha}}/\mathrm{SFR}^\mathrm{uncorr}_{\mathrm{UV}}$ for our quality ($\mathrm{flag}=2$) sample as a function of $E_\mathrm{star}(B-V)$.  SFRs are derived from $L(\mathrm{H\alpha})$ using Equation (2) of \cite{ken98} or from $L(\mathrm{UV})$ using Equation (7) of \citet{dad07}.  We see a clear correlation between this ratio and the reddening thus we can express the ratio of the observed SFRs in terms of both $E_\mathrm{star}(B-V)$ and a specific value of $f$ as follows:
\begin{eqnarray}
	\noindent \mathrm{SFR}^\mathrm{uncorr}_{\mathrm{H\alpha}}/\mathrm{SFR}^\mathrm{uncorr}_{\mathrm{UV}}=& 10^{-0.4 E_\mathrm{star}(B-V) \left( 3.33/f - 10.3 \right) }.
	\label{eq:HaUVsfr}
\end{eqnarray}
It is apparent that our data clearly falls well above the line representing the case where $f=0.44$.  While the $f=0.83$ case as determined from the Balmer decrement is more closely aligned with the observed data than the canonical value,  we find that a best-fit of Equation (\ref{eq:HaUVsfr}) to the data yields a value of $f = 0.69\pm0.02$.  This best-fit value of $f$ is specific to the present sample, the procedures adopted to estimate the SFRs and assumed extinction law.  

Based on the limitations of the current data set, we conclude that the value of $f$ likely lies between 0.69 and 0.83.  For subsequent analysis, we average the results of the two approaches to arrive at a factor $f=0.76$.  The similarity between H$\alpha$- and UV-based SFRs seen here is further checked in a  companion study that includes the far-IR data (G. Rodighiero et al., in preparation).

%Figure 3

% RESULT 3
\subsection{Star-forming Main Sequence at $z\sim1.6$}
\label{sec:MS}

\begin{figure}[t]
	\centering
	\includegraphics[height=2.4in]{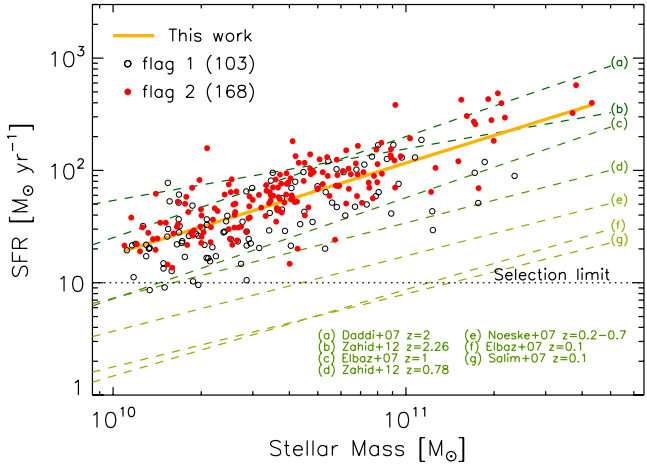} 
	\caption{SFR vs. $M_\ast$.  FMOS galaxies with flag 2 (1) are indicated by the filled (open) circles with a best-fit (yellow solid line).  Our selection limit is shown by the dotted line.  Additional lines show the MS from other published studies.}
	\label{fig:MS}
\end{figure}

We derive SFRs using Equation (2) from \citet{ken98} from extinction-corrected H$\alpha$ luminosities based on the color excess for each galaxy modified for nebular emission using $f=0.76$.  In Figure \ref{fig:MS}, SFRs are shown to cover a range of $\sim10{\rm -}500~ M_\odot~\mathrm{yr}^{-1}$ typical for star-forming galaxies at these redshifts, and exhibit a close relation with stellar mass as described in the literature as the ``main sequence''.  A fit to all H$\alpha$ detections ($\mathrm{flag}=1$ and 2) yields the following relation:     
\begin{eqnarray}
\log\mathrm{SFR} = (1.25\pm0.03) + (0.81\pm0.04)\log{\left[\frac{M_\ast}{10^{10}M_\odot} \right]}.
\label{eq:MS}
\end{eqnarray}
This relation has a very similar normalization and slope as compared to related studies at high-redshift (e.g., \citealt{dad07,wuy11}).  We recognize that the slope may be biased low due to the sensitivity limit of our survey ($\mathrm{SFR} \simeq 10~M_\odot~\mathrm{yr^{-1}}$; see Section~\ref{sec:observations}); such a bias is typical of spectroscopic samples that are more likely to measure a redshift if the SFR is above the average near the low-mass limit of the survey.  The slope is also dependent on the adopted extinction law.

We measure the width of the MS to be $\sigma_\mathrm{MS}\sim 0.22~\mathrm{dex}$, somewhat lower than typically reported values ($\sim0.25$--$0.3$ dex).  Such comparisons are not trivial since the observed width is dependent on various factors such as the redshift range considered, the imposed selection on expected H$\alpha$ flux, the impact of redshift failures, and variations in dust extinction not properly accounted for.  On the other hand, it may be that a tighter sequence is realized when employing a more accurate SFR indicator, namely, H$\alpha$ in our case.

%Figure 4

%---------------------------------------------------------------------
\section{Conclusions}
%---------------------------------------------------------------------
We are undertaking a major initiative using Subaru/FMOS to carry out a near-IR spectroscopic survey of star-forming galaxies in COSMOS at $1.4\lesssim z \lesssim 1.7$.  Our aim is to further our understanding of galaxies in an environmental context at epochs close to the peak of the SFR density.  We present the first results based on the early data to measure the SFRs based on the $\mathrm{H\alpha}$ emission line and establish the relation between the SFR and the stellar mass.  This has entailed a detailed study of the extinction properties of our sample, based both on measurements of the Balmer decrement in bins of color excess (stellar) and stellar mass, and a comparison of H$\alpha$ and UV-based SFR indicators.  Due to the complex selection of our sample and success rate of emission line detection, our results pertain to a specific subset of the overall high-redshift galaxy population and are as follows:

\begin{itemize}

\item The extinction $A_\mathrm{H\alpha}$ ranges between $\sim0.6$ and $\sim 2.5$, as measured from the Balmer decrement, and appears to exceed the corresponding values for low-redshift galaxies in SDSS at $M_\ast \gtrsim 6\times10^{10}~M_\odot$. 

\item The relation between the nebular and stellar extinction is determined to be $E_\mathrm{neb}(B-V)=E_\mathrm{star}(B-V)/f$ where $f$ is between 0.69 and 0.83.  This differs from the canonical factor ($f=0.44$) and may indicate a more uniform dust distribution in high-$z$ galaxies as compared to local galaxies.  This suggests that hot, massive stars and H{\sc ii} regions are spatially correlated more tightly in high redshift galaxies.

\item A $\mathrm{SFR}{\rm -}M_\ast$ relation based on H$\alpha$ luminosity, corrected for extinction, shows a clear correspondence between these two parameters with a slope of 0.81, a width of 0.22 dex, and a normalization ten times elevated from the local relation that is in close agreement with related high-$z$ studies.

\end{itemize}

The higher levels of extinction seen in our sample may be a consequence of both gas masses and SFRs being typically $\sim 10\times$ higher \citep[e.g.,][]{dad10}, in galaxies at $z\sim1.6$ as compared to local counterparts.  Nonetheless, we may expect some modulation of the extinction due to the lower metallicities typically seen in high-redshift galaxies, which was first noted by \cite{hay09} for a sBzK-selected sample (see also \citealt{erb06,yab12}).  Galaxy metallicities will be presented in a companion study (H.~J.~Zahid et al., in preparation).  To conclude, we highlight that further spectroscopic coverage of H$\beta$ will improve stacked spectra and average Balmer decrements, with detections for individual galaxies giving the extinction dispersion at a given mass, thus lessening biases in the current sample.

\begin{acknowledgments}
We would like to thank M. Fukugita,  T. Nozawa and V. Wild for useful discussions and K. Aoki, F. Iwamuro, and N. Tamura for their invaluable assistance and expertise regarding Subaru/FMOS.  This work has been partially supported by the Grant-in-Aid for the Scientific Research Fund under Grant Nos. 22340056: N.S., 23224005: N.A., and Program for Leading Graduate Schools ``PhD Professional: Gateway to Success in Frontier Asia'' commissioned by the Ministry of Education, Culture, Sports, Science and Technology of Japan, and INAF through the grant ``PRIN-2010''.
\end{acknowledgments}

\end{document}